\documentclass[11pt]{article}
\usepackage{amsmath}
\usepackage{amssymb}       
 \usepackage{graphicx}
\newcommand{\CP}{Chern-Pontryagin}
\newcommand{\CS}{Chern-Simons}
\newcommand{\CPR}{Clebsch parameterization}
\newcommand{\Fumn}{{F^{\mu\nu}}}
\newcommand{\Fdmn}{{F_{\mu\nu}}}
\newcommand{\Fustar}{{^*F^{\mu\nu}}}

\newcommand{\Fdab}{{F_{\alpha\beta}}}
\newcommand{\varep}{{\varepsilon^{\mu\nu\alpha\beta}}}
\newcommand{\varp}{{\varepsilon^{\mu\nu}}}
\newcommand{\vargama}{{\varepsilon^{\mu\alpha\beta\gamma}}}
\newcommand{\varijk}{{\varepsilon^{ijk}}}
\newcommand{\Fauv}{{F^{\ a}_{\mu\nu}}}
\newcommand{\Faab}{{F^{\ a}_{\alpha\beta}}}
\newcommand{\fracd}{{\frac{1}{2}}}
\newcommand{\fracdd}{{\frac{1}{4}}}
\newcommand{\gdg}{{g^{-1} dg}}

\let\Cal=\mathcal
\let\pa=\partial

\def\bz {\bar{z}}

\def\bw{{\bar w}}

\newenvironment{changemargin}[2]{%
  \begin{list}{}{%
    \setlength{\topsep}{0pt}%
    \setlength{\leftmargin}{#1}%
    \setlength{\rightmargin}{#2}%
    \setlength{\listparindent}{\parindent}%
    \setlength{\itemindent}{\parindent}%
    \setlength{\parsep}{\parskip}%
  }%
  \item[]}{\end{list}}

\title{S.~S.~Chern and \CS\ Terms\footnote{Chern Memorial, Taijin China, August 2005}}
\author{R.Jackiw\\
{\small\itshape Center for Theoretical Physics}\\
{\small\itshape Department of Physics}\\
{\small\itshape Massachusetts Institute of
Technology} \\
{\small\itshape  Cambridge, Massachusetts 02139}}
\date{\small MIT-CTP-3534}                                           

\begin{document}

\maketitle

\begin{abstract}
\begin{changemargin}{-35pt}{35pt}
\mbox{Some properties of \CS\ terms are presented and their physical utility is surveyed.}
\end{changemargin}
\end{abstract}
\newpage
\section{Meeting  S.S. Chern}
I first met Professor Chern in Durham, England a quarter century ago in the summer of 1979 at a symposium sponsored by the London Mathematical Society. The event brought together physicists and mathematicians because both discovered that after many years of separation we were again interested in common problems. This was a time when physicists realized that the axial anomaly involves the Chern-Pontryagin density, whose integral measures the topological properties of gauge fields; that the anomaly equation is a local version of the Atiyah-Singer index theorem, which in turn counts the number of zero modes in various linear elliptic equations, like the physicists' Euclidean Dirac equation \cite{50yangmills}.

I wanted to get Chern's reaction to the fact, noted by physicists, that $<^* FF >$, the axial anomaly as well as the 4-dimensional Chern-Pontryagin density, can be written as the 4-divergence of a 4-vector constructed from connections --- a quantity physicists call the anomaly current. Whereupon he informed me of the Chern-Simons secondary characteristic class, which he had put forward some years earlier \cite{chern79}. The sobriquet  ``secondary characteristic class" seems to demote that entity to a secondary class of importance. Nevertheless I was not discouraged, and with colleagues proposed using it, after renaming it simply and neutrally as the 3-dimensional Chern-Simons term \cite{Deser:1981wh}. The envisioned physical application was to dynamics  in 3-dimensional space-time, {\it i.e.}, on a plane. This suggestion was taken up by many physicists for analyzing a variety of physical, planar processes (Hall effect, high $T_c$ super conductivity, motion in presence of cosmic and other vortices). Eventually physics returned the favor to mathematics, where the Chern-Simons term describes knot invariants. 

Chern was happy that his secondary class found first class uses in physics. He thanked me for spreading the word among physicists and gave me an inscribed book, containing some of his relevant papers. 

Here I shall explore some further properties of Chern-Simons terms.

\section{Chern-Pontryagin/Simons Topological Entities}
We begin by recalling that the \CP\ densities appeared in physics when anomalous Feynman diagrams were computed. These diagrams carry vector indices, and formal arguments led us to expect that the evaluated expressions would be transverse in each index. But in fact the explicit expressions fail to be transverse, and the appropriate longitudinal part in the anomaly. In a 4-dimensional abelian gauge theory, the anomaly reads
\begin{subequations}\label{eq:01}
\begin{equation}
\Cal{A}_{(4)} = \frac{1}{2}\ \Fustar\, \Fdmn = \frac{1}{4}\ \varep \, \Fdmn\, \Fdab,
\label{eq:01a}
\end{equation}
where $\Fdmn$ is the gauge field strength (curvature). In the non-Abelian theory the expression is similar, except that the gauge fields carry a Lie algebra index $a$, that is summed.
\begin{equation}
\Cal{A}_{(4)} = \fracd\ ^*\! F^{\mu\nu\ a} \, \Fauv = \fracdd\ \varep\, \Fauv\, \Faab
\end{equation}
\label{eq:01b}
\end{subequations}
An anomaly also exists in an Abelian 2-dimensional gauge theory; it is simply
\begin{equation}
\Cal{A}_{(2)} = ^*\!\!\! F = \fracd\, \varp\, \Fdmn.
\label{eq:02}
\end{equation}
The anomalies are recognized to be densities, which upon integration over the appropriate manifold, produce the \CP\ gauge field invariant. Note that (\ref{eq:01}) and (\ref{eq:02}) are generally covariant densities, giving world scalars upon integration --- no metric tensor is needed. Because of this metric-independence, they are topological entities, independent of local, geometric properties of the manifold.

Physicist usually work on open, unbounded spaces, and the integrals are taken over these spaces. One may imagine that the integration is performed over a spherical ball,  with very large radius. The ball is bounded by its spherical surface, which passes to infinity as the radius increases without limit. Since the \CP\ entities are topological, one may expect that they can be determined by behavior at the large-distance boundary.  For this to be the case, it should be possible to represent these scalar densities as divergences of vector densities, so that by Gauss' theorem their volume integral can be cast onto the boundary surface (at infinity).

Indeed this is possible, but the field strengths (curvatures) must be expressed in terms of potentials (connections). The Abelian formula involves the familiar curl,
\begin{equation}
\Fdmn = \pa_\mu\, A_\nu - \pa_\nu \, A_\mu
\label{eq:03}
\end{equation}
while the non-Abelian expression includes a non-linear term constructed with Lie algebra structure constraints $f^{abc}$.
\begin{equation}
\Fauv = \pa_\mu\, A^a_\nu - \pa_\mu \, A^b_\nu + f^{abc}\, A^b_\mu\, A^c_\nu
\label{eq:04}
\end{equation}
Inserting (\ref{eq:03}), (\ref{eq:04}) in (\ref{eq:01}), (\ref{eq:02}) exhibits the desired result.\\[1.5ex]
\mbox{\hspace{.25in} Abelian, 4-d:}
\begin{subequations}\label{eq:05}
\begin{alignat}{2}
&\Cal{A}_{4} =\fracd \ {^*\! \Fumn}\, \Fdmn = \pa_\mu\, C^\mu_4\nonumber\\
&C^\mu_4 = \vargama A_\alpha\, \pa_\beta \, A_\gamma
\label{eq:05a}
\end{alignat}
\hspace{.25in}  non-Abelian, 4-d:
\begin{alignat}{2}
&\Cal{A}_{4} = \fracd\ {^*\! F}^{\mu\nu\, a}\, \Fauv = \pa_\mu\, C^\mu_4\\
&C^\mu_4 = \vargama \big(A^a_\gamma\, \pa_\beta\, A^a_\gamma + \frac{1}{3}\ f^{abc}\, A^a_\alpha\, A^b_\beta \, A^c_\gamma\big)
\label{eq:05b}
\end{alignat}
\end{subequations}
\hspace{.25in}Abelian, 2-d:
\begin{alignat}{2}
&\Cal{A}_2 = \fracd\, \varp\, \Fdmn = \pa_\mu \, C^\mu_2 \nonumber\\
&C^\mu_2 = \varp\, A_\nu\qquad
\label{eq:06}
\end{alignat}
The vectors $C^\mu$ whose divergence gives the anomalies $A$ are called anomaly currents or \CS\ currents.

The above is recapitulated succinctly in form notation. The anomaly or the \CP\ density is a 4-form in four dimensions and a 2-form in two dimensions. These forms are closed, and can be presented as exact forms; they are given by the exterior derivative of the \CS\ form, which is a 3-form in the former case and a 1-form in the latter.

While the \CP\ and the \CS\ forms are defined on even dimensional manifolds, one may restrict the latter, in a natural way, to one lower, odd dimensional manifold. The restriction proceeds as follows. Observe from (\ref{eq:05}) and (\ref{eq:06}) that the \CS/anomaly currents involve a free index carried by the Levi-Civita epsilon tensor. Choose a definite coordinate for that index. Because of the total anti-symmetry of the Levi-Civita tensor, the remaining indices will not repeat the chosen, external index, and therefore neither will the quantities (gauge potentials, derivatives) comprising the \CS\ current. Furthermore, if all dependence of the potentials on the selected coordinate is suppressed, we are left with the so-called \CS\ terms, defined in odd-dimensional spaces.
\begin{subequations}\label{eq:07}
\begin{eqnarray}
\mbox{Abelian, 3-d:} & CS(A) = \varijk\, (A_i\, \pa_j\, A_k)\label{eq:07a}\hspace{1.22in}\\
\mbox{non-Abelian, 3-d:}& CS(A) = \varijk \, (A^a_i\, \pa_i\, A^a_k + \frac{1}{3}\ f^{abc}\, A^a_i\, A^b_j\, A^c_k)
\label{eq:07b}
\end{eqnarray}
\end{subequations}
\vspace{-6ex}
\begin{eqnarray}
\mbox{Abelian, 1-d:} &  CS(A) = A_1\hspace{1.72in}
\label{eq:08}
\end{eqnarray}

Evidently the \CS\ terms can be integrated over 3-dimensional or 1-dimensional spaces, thereby producing world scalars without the intervention of a metric tensor. Thus we again encounter topological entities. Some of these integrals have been known in physics and mathematics for a long time, as encoding interesting properties of vector fields and gauge fields. For example, if in (\ref{eq:07a}) $A_i$ is identified with the electromagnetic vector potential, and $\varijk\, \pa_j\, A_k$ with the magnetic field $B^i$, the integral defines the  ``magnetic helicity" $\int d^3 r \, {\bf A} \cdot {\bf B}$, which measures linkage of magnetic flux lines. Alternatively, if $A_i$ is the velocity vector of a fluid $v_i$, then $\varijk\, \pa_j \, v^k \equiv \omega^i$ is the vorticity and the integral of (\ref{eq:07a}) becomes $\int d^3r\ {\bf v} \cdot {\boldsymbol \omega}$; this is the ``kinetic vorticity," which provides an obstruction to a canonical formulation of fluid mechanics \cite{example4}. When the non-Abelian \CS\ term is evaluated at a pure gauge connection $A = \gdg$ (in matrix notation), then the integrated \CS\ term involves $\int d^3r \ tr (\gdg)^3$, and evaluates the winding of the gauge function $g$. Moreover, it is known that $tr (\gdg)^3$ is a total derivative, so that the winding number integral is given by a surface term \cite{Deser:1981wh}.

\section{Chern-Simons Terms as  Total Derivatives}
The question, which this essay addresses, is whether the \CS\ terms can be expressed as total derivatives, so that their integrals over all space are given by contributions from the bounding surface.

The answer is clearly ``yes" for the 1-dimensional \CS\ term, which according to (\ref{eq:08}) is just the single function $A_1$. This can always be presented as the derivative of another quantity --- of another ``secondary" potential $\theta$,
\begin{equation}
A_4 = \pa_1\, \theta
\label{eq:09}
\end{equation}
so that $\int\nolimits^\infty_{-\infty} d x\, A_1 (x) = \theta (\infty) - \theta(-\infty)$.

Also in the 3-dimensional, Abelian case one can write the \CS\ term (\ref{eq:07a}) as a total derivative, provided the vector $A_i$ is presented in terms of further, ``secondary" potentials. 
\begin{subequations}\label{eq:10}
\begin{equation}
A_i = \pa_i \, \theta + \alpha \, \pa_i\, \beta
\label{eq:10a}
\end{equation}
The representation (\ref{eq:10a}) is called the \CPR\ of a 3-vector; it involves a ``gauge" part ($\pa_i \theta$) and two more scalars ($\alpha, \beta$), called Monge potentials. Altogether three functions appear; thus there is sufficient generality to represent the arbitrary 3-vector $A_i$. An analytic procedure for finding the \CPR\ for a given vector $A_i$ has been know since the 19th century. On the other hand, when (\ref{eq:10a}) is written in form notation
\begin{equation}
A = d \theta + \alpha \, d\, \beta
\label{eq:10b}
\end{equation}
\end{subequations}
one recognizes this as an instance of Darboux' theorem.

With $A_i$ parameterized in the \CPR\ manner, as in (\ref{eq:10}), the Abelian \CS\ term indeed becomes a total derivative.
\begin{equation}
\varijk\, A_i \, \pa_i\, A_k = \pa_i \ (\theta \varijk\, \pa_j \alpha\, \pa_k\, \beta)
\label{eq:11}
\end{equation}
With ${\bf B}\ \mbox{or}\ {\boldsymbol \omega}\ \mbox{given by}\ {\boldsymbol \nabla} \alpha  \times {\boldsymbol \nabla} \, \beta$, the magnetic helicity becomes
\begin{subequations}\label{eq:12}
\begin{equation}
\int d^3 r\ {\bf A} \cdot {\bf B} = \int d\, {\bf S} \cdot \theta {\bf B}
\label{eq:12a}
\end{equation}
and similarly for the kinetic vorticity.
\begin{equation}
\int d^3 r\ {\bf v} \cdot {\boldsymbol \omega} = \int d\, {\bf S} \cdot \theta {\boldsymbol \omega}
\label{eq:12b}
\end{equation}
\end{subequations}
Thus the volume integral of the Abelian \CS\ term is found from the surface integral of the potentials in the \CPR.

This result is important for the canonical (symplectic) formulation of Eulerian fluid mechanics. As remarked previously the kinetic vorticity provides an obstruction to a canonical formulation of that dynamical system. To make progress, the obstruction must be removed. By using the \CPR\ for the velocity places the kinetic vorticity at spatial inplicity, away from the finite regions of the 3-dimensional space, and a canonical formulation becomes possible. That is why the \CPR\ is needed in fluid mechanincs \cite{example4}.

How about the non-Abelian, 3-dimensional \CS\ term? We have already remarked that in the special case when $A_i$ is a pure gauge $g^{-1} \pa_i g$, the \CS\ term (\ref{eq:07b}) is a total derivative. We shall now show that there exists a parameterization for arbitrary (not only pure gauge) non-Abelian vectors, such that their \CS\ term is a total derivative.

\section{Mathematical Sidebar}
Before proceeding, let us reformulate our problem, and also describe work of Bott and Chern who posed and solved a related but different problem.

We know that the \CP\ entities are exterior derivatives of the \CS\ entities, as in (\ref{eq:05}) and (\ref{eq:06}).
\begin{equation}
\mbox{\CP}\ = d \mbox{(\CS)}
\label{eq:13}
\end{equation}
We have set for ourselves the problem of further demonstrating that the \CS\ quantities also are exterior derivatives of further entities.
\begin{equation}
\text{\CS}\ \stackrel{?}{=} d (\Omega)
\label{eq:14}
\end{equation}
But this also entails that 
\begin{equation}
\text{\CP} = d (\text{\CS}) = d d\, \Omega = 0
\label{eq:15}
\end{equation}
So our result can hold only when the anomaly, the \CP\ density, is absent. Thus if we work in three dimensions with a \CS\ 3-form or in one dimension with the \CS\ 1-form, the \CP\ 4-form and 2-form are absent --- they cannot be constructed. The \CS\ forms are closed because they are maximal for the considered dimensionality, and it comes as no surprise that locally exact expressions for them can be constructed.

Nevertheless, we call attention to the fact that this situation for \CS\ forms is different from the situation with \CP\ forms. The latter are closed without regard to dimensionality, whereas the former are closed for dimensional reasons.

Bott and Chern have derived a representation for the \CS\ term as a sum of (different) total derivative expressions, in the special case that the field strength (curvature) satisfies a further condition \cite{hermitian65}. To contrast and compare with our investigation we now describe their result in the 4-dimensional case. 

Bott and Chern work with the two complex coordinates that can be constructed in four dimensions.
\begin{equation}
(z, \bz) \equiv \frac{1}{\sqrt{2}} (x_1 \pm i x_2), \qquad (w, \bw) = \frac{1}{\sqrt{2}} (x_3 \pm i x_4)
\label{eq:16}
\end{equation}
They further require that the holomorphic and anti-holomorphic components of the (non-Abelian) curvature $\Fdmn$ vanish.
\begin{equation}
F_{z w} = 0 = F_{\bz \bw}
\label{eq:17}
\end{equation}
They show that then the \CP\ density takes the form
\begin{eqnarray}
\text{\CP} = d_- (\text{\CS})\nonumber\\
= d_- \, d_+\, \Omega,\hspace{1.5in}
\label{eq:18}
\end{eqnarray}
which implies that
\begin{equation*}
\text{\CS} = d_+\, \Omega + d_- \chi.
\end{equation*}
Here $d_\pm$ are the holomorphic and anti-holomorphic exterior derivatives
\begin{eqnarray}
d_+ &=& dz \frac{\partial}{\partial z} + d w \frac{\partial}{\partial w}, \label{eq:19} \\
d_{-} &=& d \bz \frac{\partial}{\partial \bz} + d \bw \frac{\partial}{\partial \bw}. \label{eq:20}
\end{eqnarray}
Thus for restricted curvatures, as in (\ref{eq:17}), the \CS\ term is a sum of terms that are exact on the holomorphic and anti-holomorphic sub-manifolds.

In contrast to the Bott-Chern result, we consider the case with no restriction on the curvature, but vanishing \CP\ density (because of dimensionality) and construct an exact (total derivative) expression for the \CS\ term. 

\section{The Result}
Our result in the non-Abelian case is not found by an analytic method, as is done for the Abelian case via the \CPR. Rather  we develop a group theoretical argument \cite{Jackiw:2000cd}. To illustrate our method, we first apply it to the Abelian case in a rederivation of the \CPR. 

To deal with the U(1) (Abelian) \CS\ term, we begin with SU(2) and consider a pure gauge connection.
\begin{equation}
A = g^{-1}\, d \, g \equiv V^a \frac{\sigma^a}{2 i}, \ g\, \epsilon \ SU(2).
\label{eq:21}
\end{equation}
The $\frac{\sigma^a}{2i}$ are the anti-Hermitian matrix generators of the Lie algebra ($\sigma^a =$ Pauli matrices).
It follows that $tr (g^{-1} d g)^3$, which is known to be a total derivative, is given by
\begin{equation}
tr (\gdg)^3 = -\frac{1}{4}\ \varepsilon_{abc}\, V^a V^b V^c = -\frac{3}{2} V^1 V^2 V^3 = \text{total derivative}.
\label{eq:22}
\end{equation}
Here $\varepsilon_{abc}$ are the $SU(2)$ structure constants. Also, because the non-Abelian connection is a pure gauge, $V^a$ obeys
\begin{equation}
d V^a = \frac{1}{2}\ \varepsilon_{abc} V^b V^c.
\label{eq:23}
\end{equation}

We define the Abelian connection, relevant to our $U(1)$ problem as
\begin{equation}
A = V^3 = tr \, i\, \sigma^3\, \gdg.
\label{eq:24}
\end{equation}
\uppercase{Note}: $A$ is not a pure gauge within $U(1)$.
It now follows that the Abelian \CS\ term satisfies the following sequence of equalities.
\begin{equation}
CS(A) = A d A = V^3 d\, V^3 = - V^1 V^2 V^3 = \frac{2}{3} \ tr (\gdg)^3
\label{eq:25}
\end{equation}
But the last term is known to be a total derivative, and this establishes that property for $CS(A)$.

Since $g$ lies in $SU(2)$, it depends on three functions, and so does $V^3$. Thus there is sufficient generality to represent an arbitrary 3-dimensional Abelian vector $A_i$.

It is instructive to see how this works explicitly. Parameterizing $g\, \epsilon \, SU(2)$ as
\begin{equation}
g = e^{\frac{\sigma^3}{2i}\, \beta} \  \, e^{\frac{\sigma^2}{2i} \gamma}\, \, e^{\frac{\sigma^3}{2i}\theta}
\label{eq:26}
\end{equation}
we find
\begin{equation*}
A = V^3 = tr \, i\, \sigma^3\, \gdg = d\, \theta + \cos \gamma \, d\, \beta
\end{equation*}
The \CPR\ is regained!

The argument for the non-Abelian \CS\ term proceeds in an analogous, but generalized manner. We seek to parameterize a connection 1-form $A^a$, belonging to the Lie algebra of $H$, whose generators are $T^a$. Consider a larger group $G$, with $H$ as a subgroup. With $g \, \epsilon\, G$ construct a pure gauge connection for  $G: \gdg$. The $H$-connection is then defined as
\begin{equation}
A^a \propto tr\, T^a\, \gdg
\label{eq:27}
\end{equation}
$A^a$ is not a pure gauge in $H$. Since an arbitrary 3-dimensional gauge potential for $H$ contains $3(\dim H)$ components, we require $\dim G > 3 (\dim H)$.

Our goal is to show that the H-\CS\ term constructed from $A^a$ coincides with the G-\CS\ term constructed from $\gdg$. The latter involves $tr (\gdg)^3$ and is known to be a total derivative. The coincidence of two then establishes that the H-\CS\ term also is a total derivative.

The desired coincidence occurs when $G/H$ is a symmetric space. In terms of the Lie algebra for $H$ and $G$, this means that the generators $T^a$ of $H, a = 1 ..., \dim H$, and the additional generators, $S^M$, $M = 1..., (\dim G - \dim H)$, which together with the $T^a$ comprise the generators of $G $, must satisfy
\begin{subequations}\label{eq:28}
\begin{alignat}{2}
&[T^a, T^b] = f_{abc}\, T^c, \label{eq:28a}\\
&[T^a, S^M] = h^{aMN}\, S^N,\label{eq:28b}\\
&[S^M, S^N] \propto h^{aMN}\, T^a.\label{eq:28c}
\end{alignat}
\end{subequations}
Eqs. (\ref{eq:28}) record the Lie algebra of $G$, with (\ref{eq:28a}) being Lie algebra of $H$ (structure constraints $f_{abc}$), with (\ref{eq:28b}) showing that the $S^M$ provide a representation for that algebra, and with (\ref{eq:28c}) giving the closure of S-generators on the T-generators.

A straightforward, but tedious, sequence of manipulations then establishes the coincidence of $G$ and $H$ \CS\ terms. To see them carried out, see the published literature \cite{Jackiw:2000cd}.

\section{Conclusion}

The 3-dimensional \CS\ term first entered physics to provide a gauge-invariant mass gap for a 3-dimensional gauge theory \cite{Deser:1981wh}. The 1-dimensional \CS\ term is related to the 2-dimensional \CS\ current $C^\mu_2 = \varp\, A_\nu$. It has recently been realized that the gauge fields in the Schwinger model --- 2-dimensional electrodynamics with massless fermions --- can be presented solely in terms of 2-dimensional topological entities: the kinetic term $\sim \Fumn \Fdmn$ is just the square of the 2-dimensional \CP\ density; the interaction with the vector current $J^\mu, J^\mu A_\mu$, is also given by $J^5_\mu C^\mu_2$ since the axial vector is dual to the vector in 2-dimensions: $J^5_\mu\, \varp = J^\nu$. Moreover, the divergence of the axial vector current is anomalous and again leads to the \CP\ density. This viewpoint on the Schwinger model suggests that it can be lifted to any even-dimensional space-time with approximate higher dimensional \CP\ densities and \CS\ currents coupled to anomalous axial vector currents. This would effect a Schwinger-model-like topological mass generation in even-dimensional space-time \cite{Dvali:2005ws}. Thus it is clear that the \CS\ term  continues to provide physicists with ideas for new physical mechanisms.

This work was supported by the U.S. Department of Energy (D.O.E) under the cooperative research agreement DE-FG02-05ER41360.

\end{document}